\documentclass[
referee, oneside, pdflatex, sn-nature]{sn-jnl}

\usepackage{graphicx,amsmath,amssymb,amsfonts,amsthm,physics,mathrsfs,physics}

\usepackage{mlmodern}

\theoremstyle{thmstyleone}%

\theoremstyle{definition}

\newcommand{\sset}[1]{\mathscr{#1}}
\newcommand{\graph}[1]{\mathscr{#1}}
\DeclareMathOperator{\sgn}{sgn}
\newcommand{\mdII}{V${}_2$}
\newcommand{\IIind}{{\text{V}_2}}

\begin{document}

\title{Symmetry-induced quantum-inspired parallelism of classical dynamic systems}

\author{\fnm{Mikhail} \sur{Erementchouk}}\email{merement@gmail.com}
\author{\fnm{Pinaki} \sur{Mazumder}}\email{pinakimazum@gmail.com}
\affil{\orgdiv{Department of Electrical Engineering and Computer Science},
  \orgname{University of Michigan}, \orgaddress{\city{Ann Arbor},
    \postcode{48104}, \state{MI}, \country{USA}}}


\abstract { Performing multiple computations within the same system,
  without spatial or temporal separation of tasks, requires encoding
  multiple data items into a well-defined physical state. The most widely
  explored mechanism for such encoding is the superposition of physical
  states representing computational states. However, superposition requires
  the system to be linear, which significantly limits the set of
  achievable operations. We show that system symmetries provide an
  alternative mechanism for encoding multiple computational states.
  Notably, this mechanism also applies to nonlinear systems and therefore
  does not impose inherent limits on computed functions.
  Using the evaluation of Boolean functions as an example, we show that a
  relaxed spin network driven by the \mdII{} model supports this
  mechanism. We relate the resulting simultaneous computations enabled by
  symmetry-induced parallelism to properties of the evaluated functions.
  We demonstrate symmetry-induced parallelism for a logical
  $\mathsf{AND}$/$\mathsf{OR}$ gate and an $N$-bit adder.  }


\maketitle

\section{Introduction}

The notion of computational parallelism is commonly associated with
computational effort performed by dedicated entities separated in space or
time. An everyday example is a multicore processor, in which different
cores---physically distinct collections of logic gates and other
components---perform their own computational tasks, either independently or
in communication with one another. A notable exception is quantum
parallelism, which employs superpositions of states in the tensor-product
space of individual qubits (see, e.g., Section~1.4.2
in~\cite{chuangQuantumComputationQuantum2000}). To a certain extent, this

approach can be mimicked in classical systems that admit superpositions of
states, for example, polarizations of the electromagnetic field. However,
to employ superpositions, the system must be linear and hence have
significantly limited computational
capabilities~\cite{millerSelfconfiguring2013, shenDeep2017,
  harrisLinear2018, feldmannParallel2021}. This is exacerbated by the fact
that reproducing the structure of the tensor product, which alleviates
limitations of linear systems in quantum computing, is challenging to scale
in classical systems~\cite{perez-garciaQuantum2015, viannaClassical2018}.

At the same time, the superposition principle can be replaced by the more
powerful principle of symmetry, which does not pose any intrinsic
limitations on dynamics and, consequently, on the system computational
capabilities. The foundational idea of symmetry-induced computational
parallelism is based on the property that, in a system with continuous
symmetries --- governed by equations of motion that are invariant under a set
of transformations --- the final state is a representative of the entire
manifold of states covered by the symmetry transformation. In turn, the
relation between the ``physical'' states, i.e., points in the system's
phase space, and the ``computational'' states represented by them need not
obey the same symmetry as dynamics. Consequently, the symmetry
transformation may traverse multiple computational states, signifying the
presence of multiple evaluations in the final state. This mechanism can be
illustrated by the following diagram
\begin{equation}\label{eq:sym-induced}
  \begin{array}{ccc}
    \boldsymbol{\psi} \left( \infty \vert \boldsymbol{\psi} \left( 0; \Sigma_{0} \right) \right)
    & \stackrel{F}{\longrightarrow}
    & f \left( \Sigma_{0} \right) \\
    \vcenter{\llap{$\mathcal{T}_{r}$}}\Big\downarrow
    &
    & \Big\downarrow\vcenter{\rlap{$\mathcal{T}_{r}$}} \\
    \boldsymbol{\psi} \left( \infty \vert \mathcal{T}_{r}[\boldsymbol{\psi} \left( 0; \Sigma_{0} \right)] \right)
    & \stackrel{F}{\longrightarrow}
      & f \left( \mathcal{T}_{r} \Sigma_{0} \right)
  \end{array}
\end{equation}
Here, $\boldsymbol{\psi}$ is the state of the dynamical system, and
$\boldsymbol{\psi} \left( \infty \vert \boldsymbol{\psi} \left( 0 \right) \right)$ denotes
the terminal state of the system evolving from the initial state
$\boldsymbol{\psi} \left( 0 \right)$. Next,
$\boldsymbol{\psi} \left( 0; \Sigma_{0} \right)$ denotes the initial state
containing the encoded computational input state $\Sigma_{0}$. The mapping
$F$ maps the state of the dynamical system to a computational state.
The
property that the evolution represents an evaluation of the function $f$ is
depicted by the upper line in the diagram, which relates
$\boldsymbol{\psi} \left( \infty \vert \boldsymbol{\psi} \left( 0; \Sigma_{0} \right) \right)$
to $f\left( \Sigma_{0} \right)$. The symmetry transformation is denoted by
$\mathcal{T}_{r}$, where $r$ is the transformation parameter. Although
diagram~\eqref{eq:sym-induced} does not presume any particular structure of
the symmetry group, within the present paper, in view of the particular
dynamical realization of symmetry-induced parallelism, we consider a
one-parameter group of transformation.

The invariance of the dynamical equations transfers the action of the
symmetry transformation to the initial state:
$\mathcal{T}_{r} \left[ \boldsymbol{\psi}\left( \infty \vert \boldsymbol{\psi}\left( 0; \Sigma_{0}
    \right) \right) \right] = \boldsymbol{\psi}\left( \infty \vert \mathcal{T}_{r}\left[
    \boldsymbol{\psi}\left( 0; \Sigma_{0} \right) \right] \right)$. In turn, the
transformation of the initial state $\boldsymbol{\psi}(0)$ may result in a
variation of the input computational state represented by
$\boldsymbol{\psi}(0)$. Finally, the right vertical arrow in
Eq.~\eqref{eq:sym-induced} defines the effective action of the symmetry
transformation on the evaluation outcome; it effectively contains the
action of the evaluated function $f$ on a (possibly nontrivial) set of
pairwise distinct input states,
$\sset{S} = \left\{ \Sigma_{0}, \Sigma_{r_{1}}, \Sigma_{r_{2}}, \ldots \right\}$, reached
through the symmetry transformations. Here,
$\Sigma_{r_1} = \mathcal{T}_{r_1}\Sigma_{0}$ is the input state reached at
$r=r_1$, $\Sigma_{r_2} = \mathcal{T}_{r_2}\Sigma_{0}$ is the input state reached at
$r = r_2$, and so on.

Diagram~\eqref{eq:sym-induced} emphasizes the property of the final
dynamical state to contain within itself the results of processing multiple
input states from the set $\sset{S}$. Of course, this diagram itself does not
guarantee that set $\sset{S}$ is nontrivial and contains any other state
besides $\Sigma_{0}$. The main result of the present paper is the demonstration
of symmetry-induced parallelism in a concrete 
dynamical system: a relaxation-based dynamical Ising machine (R-DIM) driven
by the \mdII{} model~\cite{erementchoukSelfcontained2023}.

Usually, dynamical Ising machines~\cite{mohseniIsing2022,
  kurebayashiMetrics2026}, which search for the ground state of the
classical Ising model, comprise bistable dynamical elements, so that the
machine's terminal state, with continuous dynamical variables grouping
around two values, say, $\pm 1$ in the respective units, can be immediately
interpreted as a classical spin configuration. The R-DIM abandons the
requirement of convergence to a spin-like state in favor of predictable
computational capabilities. As a result, the R-DIM's terminal states are
not necessary spin-like and may require postprocessing to extract a
feasible spin configuration. However, the structure of the terminal states
of the \mdII{} model makes such postprocessing trivial even when the terminal
state is nonbinary and there is ambiguity in the correspondence between
the machine's state and the spin configuration. As we will see, this
property plays a key role in enabling symmetry-induced parallelism.

The property of symmetry-induced parallelism is applicable in the general
information processing context. However, in the present paper, we limit
ourselves to the evaluation of Boolean functions as a representative
computational task.

\section{Results}
\label{sec:results}

\subsection{Properties enabling symmetry-induced parallelism}
\label{sec:v2-basics}

Main properties of the \mdII{} model in the context of Ising machines are
overviewed in several publications~\cite{erementchoukSelfcontained2023,
  shuklaNonbinaryDynamicalIsing2025, erementchoukRelaxationbased2025}.
Here, we only remind the features playing the key roles for the objectives
of the paper.

\begin{figure}[tb]
  \centering
  \includegraphics[width=0.95\textwidth]{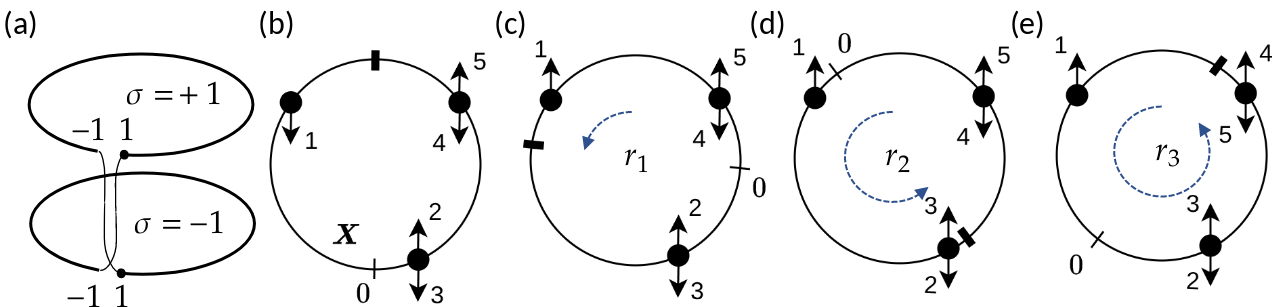}
  \caption{Relaxed spin states. (a) The phase space of relaxed spins.
    Progressing across the point of contact is accompanied with inverting
    the discrete component. (b) A terminal state of the \mdII{} model on
    $\sset{K}_{5}$ graph starting from a random initial state has the form
    of three strong clusters: two containing two relaxed spins and one
    consisting of only one. The bold dash at the top shows the
    $(-1, 1)$-boundary. Numbers near the arrows correspond to enumeration
    of the graph nodes. (c)-(e) Variations of the spin configuration with
    three counterclockwise rotations of the phase circle, or,
    alternatively, with the displacements of the origin by
    $0 < r_{1} < r_{2} < r_{3} < 2$. The dashed curved line shows the
    displacement of the $(-1, 1)$-boundary passing the strong clusters.}
  \label{fig:rs-states}
\end{figure}

The \mdII{} model governs the dynamics of $N$ dynamical variables
$\boldsymbol{\xi} = \left( \xi_{1}, \ldots, \xi_{N} \right)$ in a network described by
the weighted adjacency matrix $\widehat{A}$ with matrix elements $A_{m,n}$.
Coupling between $\xi_{m}$ is accounted by introducing new dynamical
variables, relaxed spins,
$\xi_{m} \to \psi_m = \left( \sigma_{m}, X_{m} \right)$, through the relation
$\xi_{m} = \sigma_{m} + X_{m} + kP$,
where $\sigma_{m} \in \left\{ -1, 1 \right\}$ is the discrete component of the
relaxed spin, $X_{m} \in [-1, 1)$ is the continuous component, $P$ is the
period of the coupling function and $k$ is an integer. While the last two
terms in this relation play no role in the following, it is useful to have
a context for the relaxed spins as emergent dynamical variables. Thus, the
relaxed spin dynamics is defined on the phase space of the system of
relaxed spins that has the structure shown in Fig.~\ref{fig:rs-states}(a).
When the evolution carries the phase point across the boundary of the
domain of the continuous component of the relaxed spin, the discrete
component changes sign.

A transient model state
$\boldsymbol{\psi} = \left( \boldsymbol{\sigma}, \mathbf{X} \right)$ is
characterized by an objective function that plays a role of a Lyapunov
function for deriving the equations of motion governing the evolution of
the Ising machine. The objective function can be introduced in various
ways: associating with
$\boldsymbol{\sigma}$ the energy of a classical Ising model
$\mathcal{H}(\boldsymbol{\sigma}) = 2^{-1} \sum_{m,n} A_{m,n} \sigma_m \sigma_n$ or the total weight of
edges connecting spins with opposite orientations (the weight of cut of the
network graph induced by partition described by the spin configuration)
$\mathcal{C}(\boldsymbol{\sigma}) = 4^{-1}\sum_{m,n} A_{m,n} \left( 1 - \sigma_{m} \sigma_{n} \right)$.
These approaches, while formally equivalent, emphasize different
perspective. Therefore, to simplify different discussions, we will use both
approaches throughout the paper.


The first key feature of the \mdII{} model is the
structure of the terminal states: generally, relaxed spins form groups
(\emph{strong clusters}) with coinciding continuous components (see
Theorem~1 in~\cite{erementchoukSelfcontained2023}). Figure~\ref{fig:rs-states}(b) shows
the clustered structure of the terminal state for the example of
$\sset{K}_{5}$, complete graph with five nodes.

The relaxed cut function
$\mathcal{C}_{\IIind}(\boldsymbol{\psi})$ is invariant with respect to
homogeneous translations of all continuous components:
$\mathbf{X} \to \mathbf{X} + r \mathbf{1}$, where
$\mathbf{1} = \left( 1, 1, \ldots, 1 \right)$, or, element-wise,
$X_{m} \to X_{m} + r$. Alternatively, this invariance can be interpreted as
the symmetry with respect to rotation of the phase circle (displacement of
the origin). Because of the continuity of
$\mathcal{C}_{\IIind}(\boldsymbol{\psi})$, this rotation can be extended to
arbitrary real $r$, including those that lead to strong clusters crossing
from one phase circle to another in the representation shown in
Fig.~\ref{fig:rs-states}(a) or to traverse the $(-1, 1)$-boundary in the
representation as in Fig.~\ref{fig:rs-states}(b). As a result, the symmetry
transformation leads to a variation of the discrete component of the
relaxed spin configuration. Figures~\ref{fig:rs-states}(c)-(e) show an
example of such state transformations with the original discrete component
$\boldsymbol{\sigma} = \left(-1, 1, -1, -1, 1  \right)$
[Fig.~\ref{fig:rs-states}(b)]. After the first rotation, the configuration
becomes $\boldsymbol{\sigma} = \left( 1, 1, -1, -1, 1 \right)$
[Fig.~\ref{fig:rs-states}(c)]. After the second rotation, we obtain
$\boldsymbol{\sigma} = \left( 1, -1, 1, -1, 1 \right)$
[Fig.~\ref{fig:rs-states}(d)]. Finally, the third rotation produces the
full inversion of the original state $\boldsymbol{\sigma} = \left( 1, -1, 1, 1,
  -1 \right)$ [Fig.~\ref{fig:rs-states}(exs)].

In view of these variations, the critical feature of the \mdII{} model is
that, if the system Hamiltonian is strictly quadratic in spin variables,
the correction to the relaxed cut function vanishes in the terminal state,
$\Delta\mathcal{C}_{\IIind}(\boldsymbol{\psi}) = 0$. Hence, all the discrete spin
configurations obtained while applying symmetry transformations produce the
same value of the (discrete) cut function
$\mathcal{C}(\boldsymbol{\sigma})$. This property trivializes rounding the terminal
states of the \mdII{} model: regardless of the choice of the phase
circle origin, the obtained spin configurations correspond to the same
value of the objective function (Theorem~3 in~\cite{erementchoukSelfcontained2023}). In
particular, if one configuration produces the maximum value of
$\mathcal{C}(\boldsymbol{\sigma})$, then all of them do. It is not difficult to see that
this property significantly relies on the strong clustered structure of the
terminal state. Therefore, in regularized \mdII{} models, with $\sgn(X)$ in
the right-hand-side of the equations of motion replaced by a continuous
function deviating from $\sgn(X)$ in an interval containing the origin, the
obtained spin configuration may depend on whether the displaced
$(-1, 1)$-boundary falls inside the regularization interval for some node
pairs. These situations can be consistently incorporated into the
theoretical framework with the help of the notion of \emph{weak
  clusters}~\cite{erementchoukRelaxationbased2025}. However, as will be apparent from the
following, regularizations do not present a fundamental problem, and a more
elaborated approach is excessive for the purposes of the present paper.

\subsection{\mdII-supported symmetry-induced parallelism}

The \mdII{} model is employed by representing evaluating a Boolean function
as finding the ground state (the maximum cut partition) of the respective
graph. The general principle of such
representations is
to designate some spins in the network as input bits,
$\boldsymbol{\sigma}^{(in)}$, up to the relation between binary
$\left\{ 0, 1 \right\}$ and spin $\left\{ -1, 1 \right\}$ variables, while
some spins play the role of the output bits,
$\boldsymbol{\sigma}^{(out)}$.\footnote{We adopt the convention associating spin
  values $+1$ and $-1$ with $\mathsf{True}$ and $\mathsf{False}$,
  respectively.} Consequently, the Ising Hamiltonian
$\mathcal{H}(\boldsymbol{\sigma})$ represents a vector-valued Boolean function
$f : \left\{ -1, 1 \right\}^N \to \left\{ -1, 1 \right\}^{M}$, if in the
ground state given the state of input spins $\boldsymbol{\sigma}^{(in)}$, the
state of the output spins is
$\boldsymbol{\sigma}^{(out)} = f\left( \boldsymbol{\sigma}^{(in)} \right)$. Writing
the problem of finding the Ising model ground state with fixed certain
spins as
\begin{equation}\label{eq:ground_state_restricted}
  \boldsymbol{\sigma}_* = \arg\min_{\boldsymbol{\sigma} \vert \boldsymbol{\sigma}^{(in)}}
  \mathcal{H}(\boldsymbol{\sigma}) ,
\end{equation}
the value of the Boolean function is read as the state of the output spins:
$f(\boldsymbol{\sigma}^{(in)}) = \boldsymbol{\sigma}_* \vert_{\boldsymbol{\sigma}^{(out)}}$.
Equivalently, a spin representation of a Boolean function can be regarded
as obtaining the maximum cut partition given the assignments of the input
nodes:
$\boldsymbol{\sigma}_* = \arg\max_{\boldsymbol{\sigma} \vert \boldsymbol{\sigma}^{(in)}}
\mathcal{C}(\boldsymbol{\sigma})$.

\begin{figure}[tb]
  \centering
  \includegraphics[width=0.9\textwidth]{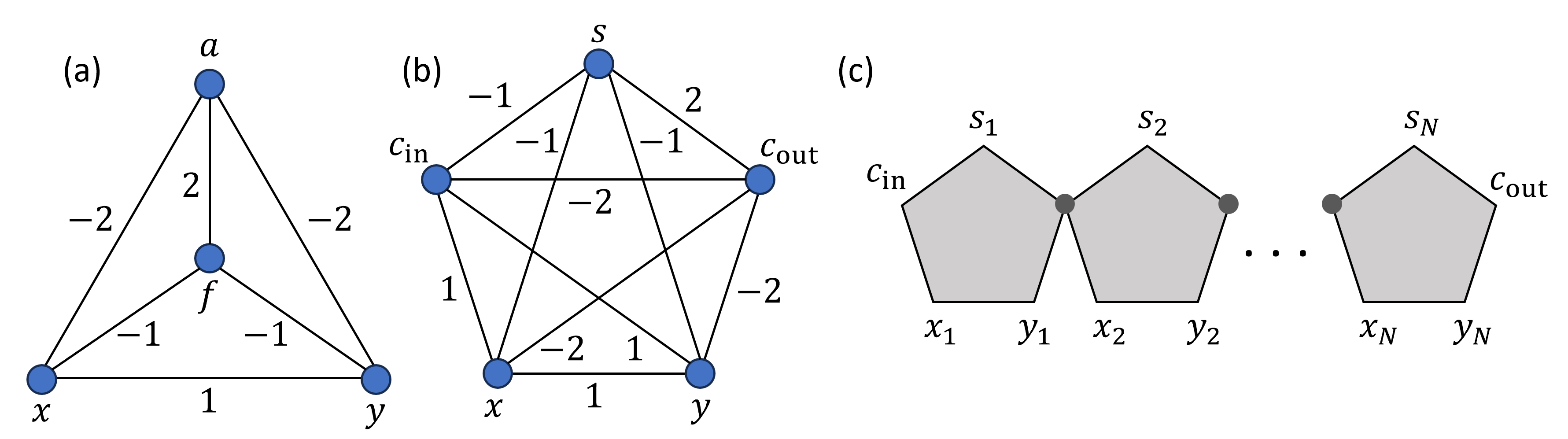}
  \caption{Spin representations of the logical gates used for illustration
    of symmetry-induced parallelism. For shown weights, values obtained
    in~\cite{tsukiyamaDesigning2025} are used. (a) $\mathsf{AND}$/$\mathsf{OR}$ gate.
    The operation regime is determined by $\sigma_f$, the value of the fixed auxiliary
    spin: $\sigma_f = 1$ and $\sigma_f = -1$ correspond to $\mathsf{AND}$ and
    $\mathsf{OR}$, respectively. (b) Full-adder, $\mathsf{FA}$. (c)
    $N$-bit adder composed of $N$ $\mathsf{FA}$'s. The carry-out spin of
    the $k$-th $\mathsf{FA}$ is identified with carry-in spin of the
    $(k+1)$-th $\mathsf{FA}$.
    }
  \label{fig:gates}
\end{figure}

Such representations were extensively studied in the
literature~\cite{whitfieldGroundstate2012a, guEncoding2012,
  caravelliLogical2020, pakhomchikConverting2022, michelSequential2025,
  tsukiyamaDesigning2025}. As we will see, symmetry-induced parallelism
poses its own set of requirements. However, as a starting point, it is
constructive to use already studied representations.
Figures~\ref{fig:gates}(a, b) show the spin representations (essentially
just complete graphs with properly chosen weights) of digital gates that we
will use to illustrate symmetry-induced parallelism: $\mathsf{AND}$ and
3-bit full adder ($\mathsf{FA}$), developed
in~\cite{tsukiyamaDesigning2025}. The respective Boolean functions can be
written in terms of their arguments as $\mathsf{AND}(x, y) = a$ and
$\mathsf{FA}\left( x, y, c_{\mathrm{in}} \right) = \langle s, c_{\mathrm{out}}\rangle$,
where $a = x \land y$, $s = x \oplus y \oplus c_{\mathrm{in}}$ is the sum bit and
$c_{\mathrm{out}} = \mathsf{MAJ}\left( x, y, c_{\mathrm{in}} \right)$ is
the carry-out bit. Here, $\oplus$ denotes $\mathsf{XOR}$ (or the sum modulo
$2$ in the $\left\{ 0, 1 \right\}$ representation for Boolean variables),
and $\mathsf{MAJ}$ is the majority function (returns $\mathsf{True}$, if
and only if the most arguments are $\mathsf{True}$). Taking into account
these notations, we write the Hamiltonians describing the spin
representations of these gates as
\begin{equation}\label{eq:spin-gates}
  \begin{split}
    \mathcal{H}_{\mathsf{AND}}(\boldsymbol{\sigma}) & =
                                       \sigma_x \sigma_y - 2\sigma_x \sigma_a - 2\sigma_y \sigma_a - \sigma_x
                                       \sigma_f - \sigma_y \sigma_f + 2\sigma_a \sigma_f, \\ 
    \mathcal{H}_{\mathsf{FA}}(\boldsymbol{\sigma}) & =
                                      2\sigma_{\mathrm{o}} \sigma_s -
                                      2\sigma_{\mathrm{o}} \sigma_x - 2\sigma_{\mathrm{o}}
                                      \sigma_y - 2\sigma_{\mathrm{o}} \sigma_{\mathrm{i}}
    \\
                                     & -
                                       \sigma_s \sigma_x - \sigma_s \sigma_y - \sigma_s\sigma_{\mathrm{i}}
                                       + \sigma_x \sigma_y +
                                       \sigma_x \sigma_{\mathrm{i}} + \sigma_y\sigma_{\mathrm{i}},
  \end{split}
\end{equation}
where $\sigma_f \equiv 1$ is a fixed auxiliary spin, and
$\sigma_{\mathrm{i}}$ and $\sigma_{\mathrm{o}}$ are spins representing carry-in and
carry-out, respectively.

The fixed auxiliary spin is introduced in the spin representations of
$\mathsf{AND}$ to ensure that $\mathcal{H}_{\mathsf{AND}}(\boldsymbol{\sigma})$ is
strictly quadratic in spin variables. This reflects the fact that the truth
table of $\mathsf{AND}$ is not invariant with respect to inversion of all
spins (such an inversion yields $\mathsf{OR}$ function), while
$\mathsf{FA}$ does have this symmetry.

\begin{figure}[tb]
  \centering
  \includegraphics[width=0.9\textwidth]{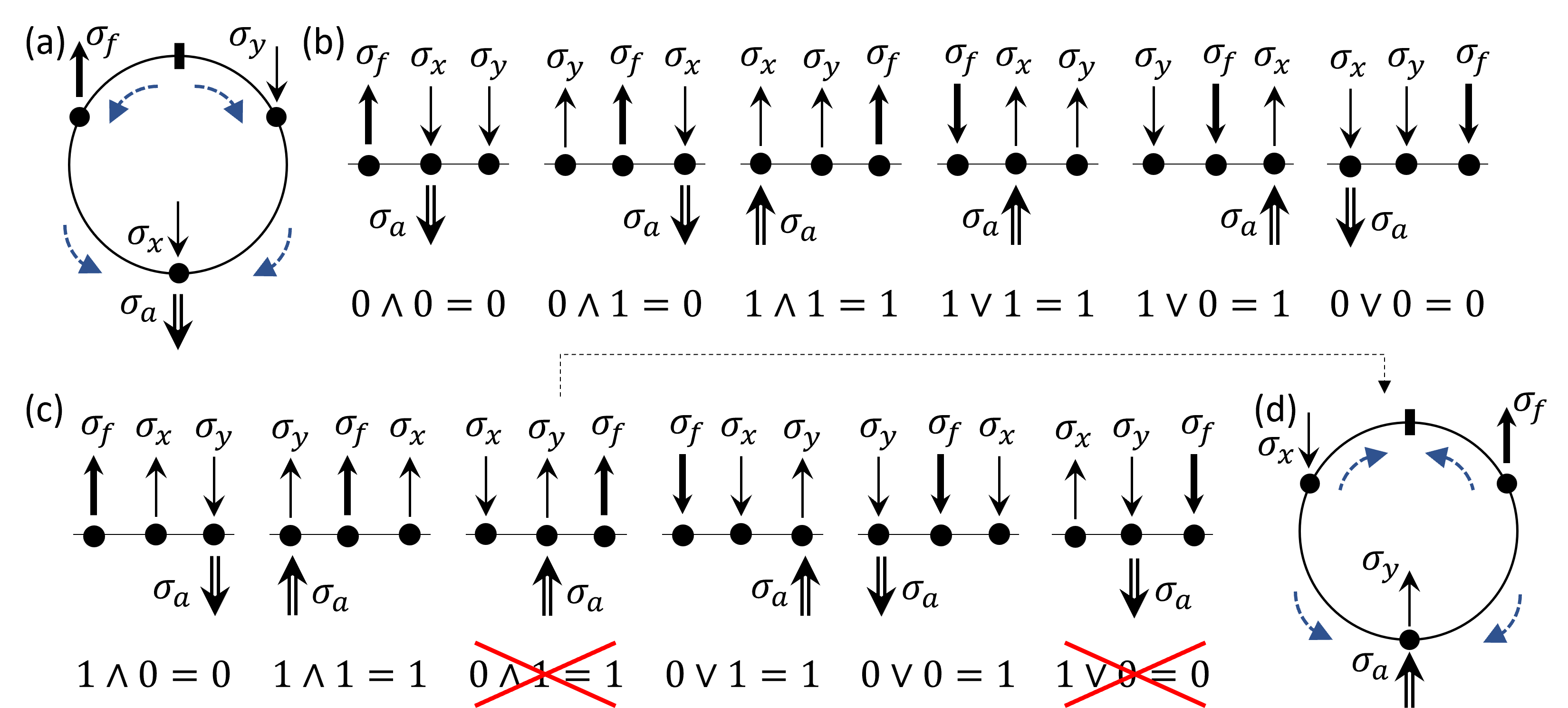}
  \caption{Symmetry-induced parallelism in the $\mathsf{AND}$/$\mathsf{OR}$
    gate. The bold and split arrows show the auxiliary spin and the output
    relaxed spin, respectively. Bold dots indicate clusters in the stable
    equilibrium state: two clusters are trivial, and one cluster contains
    two relaxed spins. (a) An example of arrangement of the input spins and
    the solution delivered by the \mdII{} model corresponding to the
    starting state in chain shown in (b). Curved dashed lines indicate
    $\dot{X}_a$ of the relaxed spin $\psi_a = \left( \sigma_a, X_a \right)$ placed
    inside the respective intervals with $\sigma_a = 1$. (b) An example of a
    maximal chain encoded in the relaxed spin state shown in (a). (c) The
    only chain inconsistent with the gate functionality. (d) The stability
    diagram of the state inconsistent with the gate functionality.}
  \label{fig:and-chaines}
\end{figure}

We start with considering in detail symmetry-induced parallelism for
the example of the $\mathsf{AND}$ gate, and then demonstrate it for the
$N$-bit adder. Symmetry-induced parallelism is enabled by the clustered
structure of the \mdII{} terminal states. This feature is engaged by
employing the continuous degree of freedom of the relaxed spins for storing
information. The spins representing the input arguments are set at fixed
displaced positions on the phase circle, as illustrated by
Fig.~\ref{fig:and-chaines}(a). If the relaxed spin system ends up in a
proper state, for example, as shown in Fig.~\ref{fig:and-chaines}(a) by the
split arrow, the resultant state of relaxed spins, including the auxiliary
and input ones, represents a series of binary states corresponding to
evaluating $\mathsf{AND}$ and $\mathsf{OR}$ for various input arguments.
Thus, in the case shown in Fig.~\ref{fig:and-chaines}(a), the single
evaluation of the relaxed spin terminal state produced the results of six
computations. These results can be read-off by applying the respective
symmetry transformation, for instance, by rotating the phase circle, as
illustrated by Fig.~\ref{fig:and-chaines}(b). It is worth noting that the
transformations required for extracting the results are fully determined by
the arrangements of the fixed spins on the phase circle.

An arrangement of relaxed spins along the phase circle defines the order in
traversing various function arguments. A convenient framework for
describing transformations occurring during single rotation of the phase
circle is that of partially ordered sets (standard background on posets,
chains, antichains, and other related concepts can be found in,
e.g.,~\cite{mazurCombinatorics2009}). The power set of the set of indices
enumerating relaxed spins representing the input parameters is ordered by
inclusion. A \emph{chain}
$\sset{J} : \emptyset \prec \left\{ m_1 \right\} \prec \ldots$ defines a sequence of spin-flip
transformations. Applying this sequence to a binary state produces a chain
of binary states. Thus, a terminal state of the \mdII{} model plays the
role of the basepoint of the chain of states yielded by the symmetry
transformation. Notably, this sequence of states,
$\boldsymbol{\sigma}(0) \prec \boldsymbol{\sigma}(1) \prec \ldots$, forms chains while restricted
to both spins representing the function arguments,
$\boldsymbol{\sigma}^{(in)}(0) \prec \boldsymbol{\sigma}^{(in)}(1) \prec \ldots$, and spins
representing the function output,
$\boldsymbol{\sigma}^{(out)}(0) \prec \boldsymbol{\sigma}^{(out)}(1) \prec \ldots$.

The same property can be approached from the perspective of the Boolean
functions, $f : \left\{ -1, 1 \right\}^N \to \left\{ -1, 1 \right\}^M$,
itself. A chain of bit-flip transformations
$\sset{J}^{(in)} : \emptyset = J_0 \subset J_1 \subset \ldots$ applied to the vector of the function
arguments $\boldsymbol{\sigma}^{(in)}$ produces a chain of arguments
$\sset{C}^{(in)} : \boldsymbol{\sigma}^{(in)} = \boldsymbol{\sigma}^{(in)}(0) \prec
\boldsymbol{\sigma}^{(in)}(1) \prec \ldots$. We say that function $f$ is
$\sset{C}^{(in)}$-isotone~\cite{rudeanuLocal1975} if
$f\left( \boldsymbol{\sigma}^{(in)}(0) \right) \preceq f\left(
  \boldsymbol{\sigma}^{(in)}(1) \right) \preceq \ldots$, with the order transported from
the respective bit-flip chain $\sset{J}^{(out)}$ in
$\left\{ -1, 1 \right\}^M$.

These two perspectives are connected by observing that during the symmetry
transformation, each spin inverts at most once. Then, there exists such an
association between components of $\boldsymbol{\sigma}^{(out)}$ and
$\boldsymbol{\sigma}^{(in)}$ (we will call it \emph{consistent} with $f$) that
the symmetry transformation produces both bit-flip chains,
$\sset{C}^{(in)}$ and $\sset{C}^{(out)}$, so that
$\boldsymbol{\sigma}^{(out)}(0) = f\left( \boldsymbol{\sigma}^{(in)}(0) \right)$,
$\boldsymbol{\sigma}^{(out)}(1) = f\left( \boldsymbol{\sigma}^{(in)}(1) \right)$, and
so on. Figure~\ref{fig:and-chaines}(a) shows an example of such an
association for the $\mathsf{AND}$/$\mathsf{OR}$ function.

Using this framework, we show that, at least for simple functions, spin
networks driven by the \mdII{} model enables symmetry-induced parallelism.
To this end, we consider a spin network implementing a
(scalar) Boolean function
$f: \left\{ -1, 1 \right\}^N \to \left\{ -1, 1 \right\}$, where $N$ includes
the auxiliary spins, if needed. The main simplifying assumption is that the
network implements the function without involving additional internal
relaxed spins, so that the only dynamically variable spin is the one
representing the function output. 
Next, let the relaxed spins representing
function arguments are arranged on the phase circle in such a way that the
symmetry transformation produces a bit-flip chain $\sset{C}$, and function
$f$ is $\sset{C}$-isotone. Then, the state of the output relaxed spin
consistent with $f$ is the only stable equilibrium of the network.

To show this, we consider the phase circle as the interval
$\left[ -1, 1 \right]$ and arrange input relaxed spins $\psi_m$, with
$1 \leq m \leq N$, along the phase circle with pair-wise different values of the
continuous component, similarly to Fig.~\ref{fig:and-chaines}(a, d).
Without loss of generality, we assume that input relaxed spins are
enumerated in the order of traversing the phase circle counterclockwise:
$X_m \leq X_n$, if $m < n$. Next, we consider the sign of $\dot{X}_a$ when the
dynamical relaxed spin $\psi_a$ is placed inside different intervals formed by
the boundaries $\left\{ \pm 1 \right\}$ and positions of the fixed relaxed
spins. Let there be $S \leq N +1$ such intervals. In the outmost right, $S$-th,
interval, $X_a = X^{(S)}$ with $X_N < X^{(S)} < 1$, we have
\begin{equation}\label{eq:Xa_dot_right}
  \sgn\left[ \dot{X}_a \left( X^{(S)} \right)  \right] = 
  \sigma_a \sgn \left[ F\left( \boldsymbol{\sigma}^{(in)} \right)  \right],
\end{equation}
where
$F\left( \boldsymbol{\sigma}^{(in)} \right) = \sum_{m} A_{a, m} \sigma_m^{(in)}$, with
$\boldsymbol{\sigma}^{(in)} = \boldsymbol{\sigma}^{(in)}(0)$, depends only on the
spins representing the function arguments at the original orientation of
the phase circle. The necessary condition of maximum cut coincides with
$\sigma_a F\left( \boldsymbol{\sigma}^{(in)} \right) \leq 0$. This inequality signifies
that inverting $\sigma_a$ does not increase the total weight of cut edges
incident to $a$. Since maximizing cut determines $\sigma_a$ unambiguously, we
presume that $\sigma_a F\left( \boldsymbol{\sigma}^{(in)} \right)$ is strictly
negative. In other words, $\sgn \left[ F(\boldsymbol{\sigma}) \right]$ is
negation of the represented function,
$\sgn \left[ F(\boldsymbol{\sigma}) \right] = \neg f(\boldsymbol{\sigma})$.

Thus, when the spin configuration
$\left( \boldsymbol{\sigma}^{(in)}, \sigma_a\right)$ satisfies the function truth
table (yields the maximum cut), the free relaxed spin $\psi_a$ goes away from
the right boundary of $\left[ -1, 1 \right]$. Since the equations of motion
are antisymmetric with respect on inversion of the phase circle, in the
outmost left interval, $X_a = X^{(1)}$ with $-1 < X^{(1)} < X_1$, we have
$\sgn\left[ \dot{X}_a\left( X^{(1)} \right) \right] = - \sgn\left[
  \dot{X}_a\left( X^{(S)} \right) \right]$, and $\psi_a$ goes away from the
left boundary, as well. Consequently, the state yielding the output of the
Boolean function is dynamically stable and attracting, with $X_a(t)$
terminating inside the interval $\left[ X_1, X_N \right]$.

Next, we notice that the sign of $\dot{X}_a$ in an
``internal'' interval is the same as when this interval becomes outmost
right after the symmetry transformation:
\begin{equation}\label{eq:Xa_dot_next_right}
  \sgn\left[ \dot{X}_a \left( X^{(S-k)} \right)  \right] = 
  \sigma_a \sgn \left[ F\left( \boldsymbol{\sigma}^{(in)}(k) \right)  \right].
\end{equation}
This follows from the symmetry of the equations of motion, or can be
directly checked by tracing how the sign of the right-hand-side of the
equations of motion changes while crossing from one interval to another.
Hence, $\sgn \left[ F \left( \boldsymbol{\sigma} \right) \right]$ is
$\sset{C}$-isotone and changes sign only once while traversing the
intervals. The boundary between the intervals where the sign changes is the
stable equilibrium of the \mdII{} model, and this is the sole stable
equilibrium.

When the represented function is not isotone along the bit-flip chain
induced by the symmetry transformation, we will call the flip chain
\emph{broken}. For example, for the $\mathsf{AND}$/$\mathsf{OR}$ gate, such
case is shown in Fig.~\ref{fig:and-chaines}(c). It must be noted that out
of $11$ different nontrivial chains, that is chains that are not related to
each other by the symmetries of $\mathsf{AND}$ and $\mathsf{OR}$ functions
and the phase circle, this is the only broken full chain. All other chains
are fully consistent with the gate truth table.

It must be noted that, for given function $f$, being isotone along a
bit-flip chain is the property of the chain. Even if the function is not
isotone along a saturated chain (with only one bit inverted at each step
along the chain), the function may be isotone along coarser chains. As an
ultimate example, any function is isotone along bit-flip chains with less
than three elements.

Thus, for \emph{any} (vectorial) Boolean function, symmetry-induced
parallelism, supported by the uniform rotation of the relaxed spin phase
circle, enables performing, at least, two simultaneous computations,
in view of a single rotation of the phase circle, or four, in view of full
double rotation. In other words, for any Boolean function,
$f(\boldsymbol{\sigma})$, separating the arguments into two groups,
$\boldsymbol{\sigma} = \left( \boldsymbol{\sigma}^{(A)}, \boldsymbol{\sigma}^{(B)}
\right)$, enables chain $f\left( \boldsymbol{\sigma}^{(A)}, \boldsymbol{\sigma}^{(B)}
\right) \to f\left( \boldsymbol{\sigma}^{(A)}, -\boldsymbol{\sigma}^{(B)}
\right) \to f\left( -\boldsymbol{\sigma}^{(A)}, -\boldsymbol{\sigma}^{(B)}
\right) \to f\left( -\boldsymbol{\sigma}^{(A)}, \boldsymbol{\sigma}^{(B)}
\right)$. 

For the relaxed spin networks this means that broken chains can be avoided
if the number of varied parameters is reduced by clustering some of them.
For the $\mathsf{AND}$/$\mathsf{OR}$ gate, assigning to the relaxed spins
$\psi_x$ and $\psi_y$ the same value of the continuous component makes them
revert simultaneously with the rotation of the phase circle thus limiting
the maximum number of variations to two. The same approach to reducing the
number of variable arguments will be used below for the adder.

Broken chains may appear contradicting to the key property of the \mdII{}
model's terminal states. However, since relaxed spins encoding multiple
sets of arguments are fixed during the evolution, the relaxation-related
correction to the cut function,
$\Delta \mathcal{C}_{\IIind}(\boldsymbol{\psi})$ [Eq.~\eqref{eq:v2-cut-correction}], does not
necessarily vanish in an equilibrium state. While
$\Delta \mathcal{C}_{\IIind}(\boldsymbol{\psi})$ is independent of $X_a$ in an equilibrium
state, it does depend on continuous components of the fixed relaxed spins.
Consequently, the symmetry transformation, while preserving the stability
of the terminal state with respect to small perturbations, as illustrated
by Fig.~\ref{fig:and-chaines}(d), may lead to variations of the discrete
cut $\mathcal{C}(\boldsymbol{\sigma})$ (as typical for relaxations). The broken chain
corresponds to the case where the rotation of the phase circle results in a
spin configuration yielding a smaller cut.

Following essentially the same argument about the structure of the Boolean
function and the model dynamical properties, one can see that, dynamically,
broken chains manifest in (and, in fact, are equivalent to) the existence
of multiple stable equilibria. For example, for the first state in
Fig.~\ref{fig:and-chaines}(c), another stable equilibrium is when $\psi_a$
clusters with $\psi_f$. Settling in different stable equilibria determines
different variations of the output parameter with the symmetry
transformation, which cannot be consistent with the function output.

\subsection{$N$-bit adder}

To demonstrate symmetry-induced parallelism in a more complex
setup, we consider $N$-bit ripple-carry adder, constructed of $N$ full
adders $\mathsf{FA}$, as shown in Fig.~\ref{fig:gates}(b, c), by
identifying the spin representing the carry-out bit in the $k$-th
$\mathsf{FA}$ gate with the spin representing the carry-in bit in the
$(k + 1)$-th $\mathsf{FA}$ gate. Alternatively, these bits could be
represented by different nodes connected by an edge with a negative weight.

It is worth noting that constructing spin representations of complex
Boolean functions out of more elementary, say, logical gates, poses little
constraints on spin representations of individual sub-functions. It is
sufficient that the cut values obtained whenever the gate equation is
satisfied,
$\boldsymbol{\sigma}^{(out)} = f_i\left( \boldsymbol{\sigma}^{(in)} \right)$, where
$i$ is the gate number, are the same
$\max_{\boldsymbol{\sigma} | \boldsymbol{\sigma}^{(in)}} \mathcal{C}_i(\boldsymbol{\sigma}) = C$ for
all gates, while the cut values corresponding to unsatisfied gate equations
are strictly smaller. Then, for any consistent circuit, that is such that
there exists an assignment of Boolean values to all wires satisfying
simultaneously every gate equation, using the spin representations of
individual gates will produce a valid spin representation $\graph{G}$ of
the whole circuit. Indeed, let us consider a maximum cut of this graph
given the values of input parameters
$\overline{C}_{\graph{G}} = \mathcal{C}_{\graph{G}}(\boldsymbol{\sigma}_*)$. Obviously,
the cut of any induced subgraph evaluated for the respective subset of
$\boldsymbol{\sigma}_*$ cannot exceed the maximum cut of this subgraph. Since
the spin representations of individual gates do not share edges, we have
$\overline{C}_{\graph{G}} \leq N C$, where $N$ is number of gates in the
circuit. On the other hand, since the circuit is consistent, one can assign
the input and output spins of each gate according to the Boolean states of
the circuit wires given the input. Such an assignment will maximize the cut
for each gate representing graph and, hence,
$\overline{C}_{\graph{G}} \geq N C$. Thus, we have
$\overline{C}_{\graph{G}} = N C$, and $\boldsymbol{\sigma}_*$ is the spin
assignment satisfying all gate equations in the circuit with the circuit
output spins corresponding to the value of the represented Boolean function
given its input.

However, the existence of broken chains prevents translating this argument
straightforwardly to the property of symmetry-induced parallelism. As the
discussion in the previous section suggests, we reduce the number of
variable parameters by clustering the input bits $x_i$ and $y_i$. In a
single $\mathsf{FA}$, this corresponds to having two arguments,
$c_{\mathrm{in}}$ and $(x, y)$, that can be varied by the symmetry
transformation and two output parameters, $c_{\mathrm{out}}$ and $s$. Of
course, in a single $\mathsf{FA}$, other pairings are possible, however, in
the ripple-carry adder, the relaxed spins representing the carry-in bits in
the $k$-th $\mathsf{FA}$'s, with $k > 1$, are determined dynamically and
should not be tied to the input bits. Thus, the number of simultaneous
computations (the length of a maximum chain) obtained through 
symmetry-induced parallelism of the $N$-bit adder is limited from above by
$2N + 2$.

\begin{figure}[tb]
  \centering
  \includegraphics[width=0.9\textwidth]{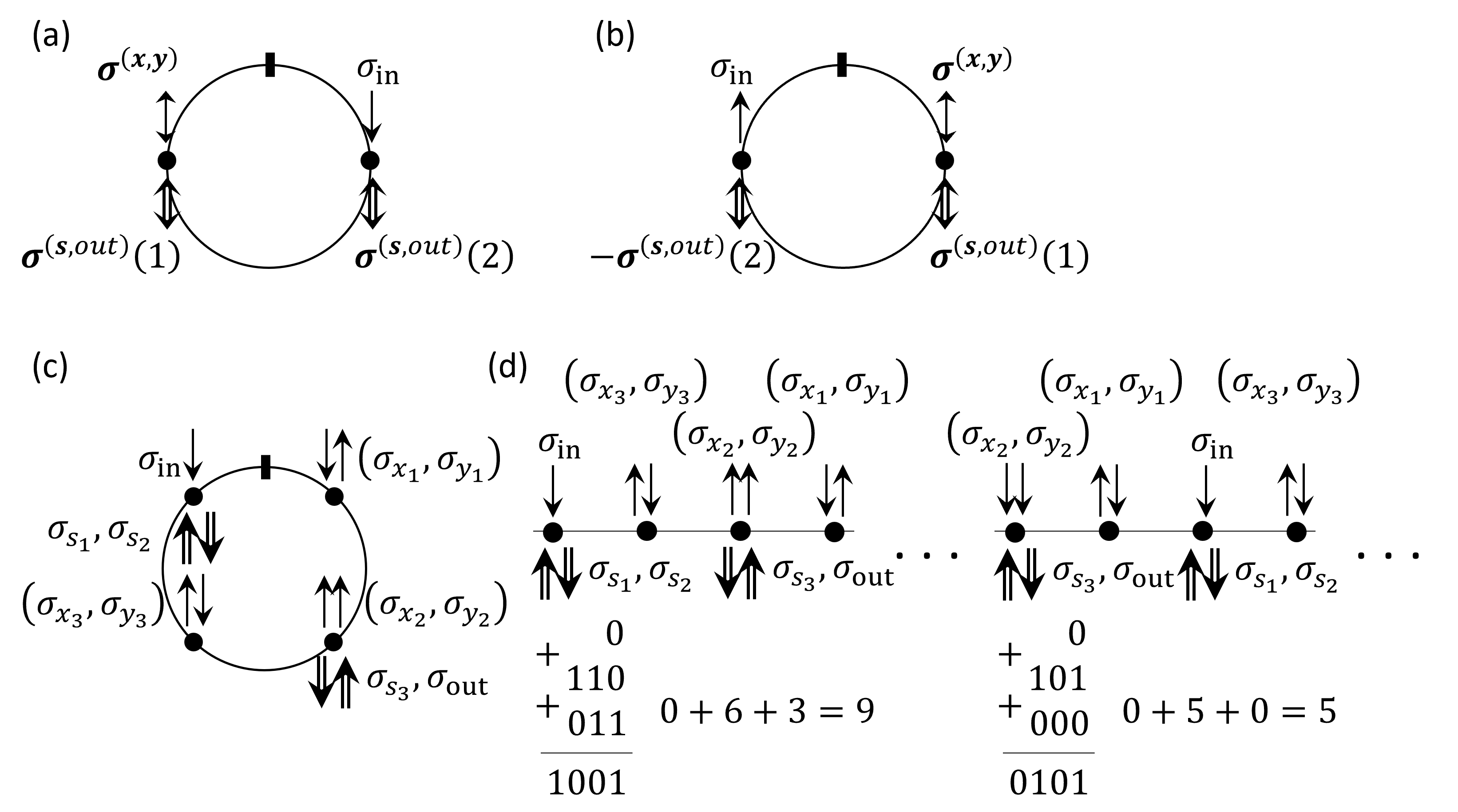}
  \caption{Structure of states representing multiple outputs for $N$-bit
    adder. (a, b) The carry-select effect reproduced by 
    symmetry-induced parallelism:
    $\boldsymbol{\sigma}^{(\mathbf{x}, \mathbf{y})}$ shows that all input
    relaxed spins are clustered together,
    $\boldsymbol{\sigma}^{(\mathbf{s}, out)}(1)$ and
    $\boldsymbol{\sigma}^{(\mathbf{s}, out)}(2)$ show the output spins
    (including the carry-out spin) clustered with the input and carry-in
    spins, respectively. (a) The carry-in bit is reset,
    $c_{\mathrm{in}} = 0$, (b) the carry-in bit is set,
    $c_{\mathrm{in}} = 1$. (c) An example of a state at the start of a
    maximal chain for a $3$-bit adder. (d) The first and the third states in
    the chain starting from the state shown in (c):
    $0 + 6 + 3 = 9$, $0 + 7 + 2 = 9$,
    $0 + 5 + 0 = 5$, $0 + 4 + 1 = 5$, $1 + 4 + 1 = 6$, $1 + 0 + 5 = 6$,
    $1 + 2 + 7 = 10$, $1 + 6 + 3 = 10$.}
  \label{fig:add-structure}
\end{figure}

Figure~\ref{fig:add-structure} illustrates the internal structure of 
symmetry-induced parallelism for $N$-bit adders.
Figures~\ref{fig:add-structure}(a, b) show how this parallelism reproduces
the carry-select effect~\cite{parhamiComputer2010}. This is a consequence of the
discussed above fact that any Boolean function is isotonic along chains
with two elements. In this case, the input relaxed spins are clustered
together, while the relaxed spin representing the carry-in bit is displaced
on the phase circle. The spins representing the output sum and the
carry-out bit are dynamically distributed between these two clusters. This
is depicted schematically in Fig.~\ref{fig:add-structure}(a,b) by two
vectors: $\boldsymbol{\sigma}^{(out)}(1)$ corresponding to the relaxed spins
that cluster with the input spins, and $\boldsymbol{\sigma}^{(out)}(2)$
representing those clustering with the carry-in spin. Depending on the
actual value of the carry-in bit, the sum is read either directly from the
terminal state ($c_{\mathrm{in}} = 0$) or after rotating the phase circle
clockwise by $\pi/2$ ($c_{\mathrm{in}} = 1$). In the latter case, the spins
collected in $\boldsymbol{\sigma}^{(out)}(2)$ change their sign.

From the computer arithmetic perspective, allowing for the carry-in bits in
individual $\mathsf{FA}$'s to settle dynamically can be regarded as $N$-bit
adder implemented as $N$ $1$-bit carry-select adders. In this regard, it
must be emphasized that in conventional carry-select adders each
computation, assuming a particular value of the carry-in bit, is performed
in a dedicated block. Owing to symmetry-induced parallelism, in the
relaxed spin realization of $N$-bit adder, all computations are done within
the same system.

Figure~\ref{fig:add-structure} shows the first state of a chain for a
$3$-bit adder. The state represents ``$0 + 6 + 3 = 9$'', where $0$
corresponds to the carry-in bit. It should be noted that the order of
arrangement of the variable inputs matters. Different arrangements may
produce different chains that coincide only at $0$, $2\pi$ (full inversion of
the discrete component of the relaxed spin), and $4\pi$ (return to the
original state) angles of rotation of the phase circle.

\begin{figure}[tb]
  \centering
  \includegraphics[width=0.975\textwidth]{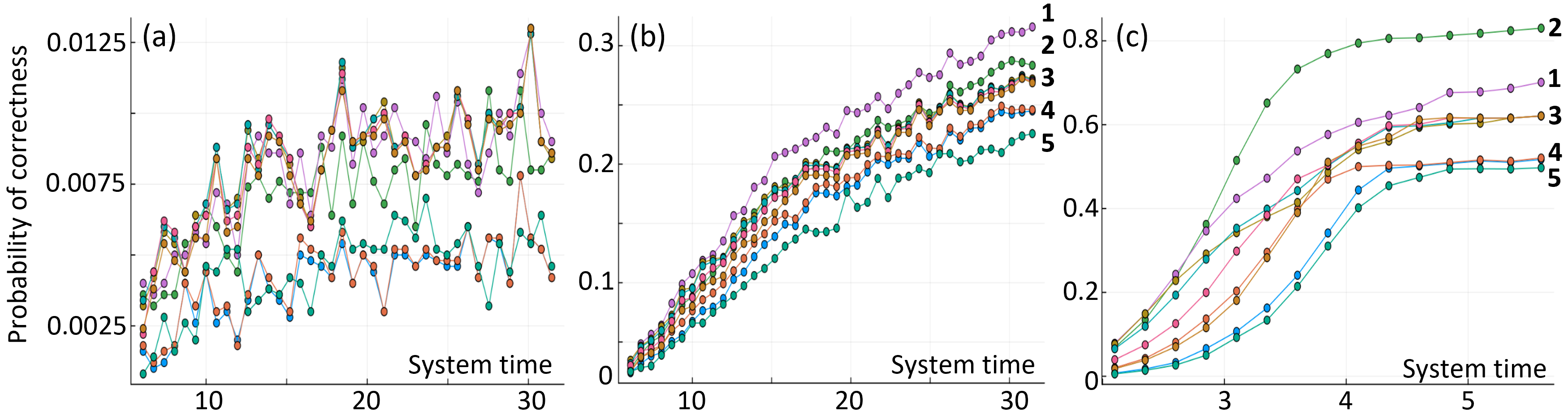}
  \caption{An example of the probability to obtain correct results for
    individual branches of adding two $32$-bit numbers on the duration of
    evolution in units of the model equations of motion
    [Eq.~\eqref{eq:v2-eqm}]. (a) Concurrent $N$-bit adder with the weights
    shown in Fig.~\ref{fig:gates}(b)~\cite{tsukiyamaDesigning2025}. (b) Concurrent
    $N$-bit adder with the weights updated as described in the text. The
    numbers in braces in (b) and (c) designate groups of solutions with
    close probabilities. (c) Sequential $N$-bit adder with the updated
    weights of individual $\mathsf{FA}$'s. The dependence on the duration
    of running of each individual $\mathsf{FA}$ is shown.}
  \label{fig:adder-probabilities}
\end{figure}

While these considerations demonstrate the principal possibility of the
system-induced parallelism in a spin network driven by the \mdII{} model,
they leave open the question of whether the network converges to the
required state starting from a generic initial state. As
Fig.~\ref{fig:adder-probabilities} demonstrates, during the evolution,
the system of relaxed spins may encounter dynamical bottlenecks that may
lead the system to a metastable state.

Figure~\ref{fig:adder-probabilities} shows the probability of obtaining
correct results for the example of adding two $32$-bit numbers and the
carry-in bit $0 + 3411433493 + 2079581652$. The parallel computations were
encoded by displacing relaxed spins $\langle17, 3, 25, 0, 8, 30, 12, 21 \rangle$, with
$0$ standing for the relaxed spin representing the carry-in bit, as seen
while traversing the phase circle from $X = 1$ to $X = -1$. Thus, this
arrangement of input relaxed spins encoded the following branches
\begin{equation}\label{eq:add-branches-example}
  \begin{split}
    \left\{ \emptyset  \right\} & \qquad 0 + 3411433493 + 2079581652 = 5491015145, \\
    \left\{ 17 \right\} & \qquad 0 + 3411499029 + 2079516116 = 5491015145, \\
    \left\{ 3  \right\} & \qquad 0 + 3411499025 + 2079516112 = 5491015137, \\
    \left\{ 25 \right\} & \qquad 0 + 3394721809 + 2062738896 = 5457460705, \\
    \left\{ 0  \right\} & \qquad 1 + 3394721809 + 2062738896 = 5457460706, \\
    \left\{ 8  \right\} & \qquad 1 + 3394721937 + 2062738768 = 5457460706, \\
    \left\{ 30 \right\} & \qquad 1 + 3931592849 + 1525867856 = 5457460706, \\
    \left\{ 12 \right\} & \qquad 1 + 3931590801 + 1525869904 = 5457460706, \\
    \left\{ 21 \right\} & \qquad 1 + 3930542225 + 1524821328 = 5455363554, 
  \end{split}
\end{equation}
where $\left\{ \emptyset \right\} $ corresponds to the terminal state as seen at
the original orientation of the phase circle, $\left\{ 17 \right\}$ denotes
the inversion of the $17$-th pair of bits in the input numbers,
$\left\{ 3 \right\}$ corresponds to additional inversion of the $3$-rd
pair, and so on.

The characteristic feature of results shown in
Eq.~\eqref{eq:add-branches-example} is the presence of several groups with
the same result. These groups correspond to inverting pairs $(0,1)$ that
change arguments but preserve the sum. These groups translate to groups with
close probabilities of obtaining correct results as evidenced by groups
marked in Figs.~\ref{fig:adder-probabilities}(b, c):
$\boldsymbol{1} \to \left\{ 25 \right\}$,
$\boldsymbol{2} \to \left\{ 3 \right\}$,
$\boldsymbol{3} \to \left\{ 0, 8, 30, 12 \right\}$,
$\boldsymbol{4} \to \left\{ \emptyset, 17 \right\}$,
$\boldsymbol{5} \to \left\{ 21 \right\}$, with the branches identified
according to the additional bit inversions, as shown in the first column
in Eq.~\eqref{eq:add-branches-example}.

The weights shown in Fig.~\ref{fig:gates}(b) admit a variety of bottlenecks
resulting in a poor convergence of the $N$-bit adder as evidenced by
Fig.~\ref{fig:adder-probabilities}(a). Some bottlenecks can be related to
the linear dependence of weights in Fig.~\ref{fig:gates}(b) over
$\left\{ -1, 1 \right\}$. As a simple ``fix'', we removed some linear
dependencies by increasing the magnitude of coupling in pairs
$\left( x, c_{\textrm{out}} \right)$ and
$\left( y, c_{\textrm{out}} \right)$ by 10 percent, and by scaling the
weights in the $k$-th $\mathsf{FA}$ by
$\sqrt{p_k} / \lfloor\sqrt{p_k}\rfloor$, where $\left\{ p_k \right\}$, with
$k = 1, \ldots, N$, is a set of random distinct prime numbers (any set of
numbers linearly independent over rationals suffices).
Figure~\ref{fig:adder-probabilities}(b) confirms that eliminating linear
dependencies drastically improves the probability of converging to correct
solution. To demonstrate the performance of a cascaded circuit in the setup
with eliminated mutual impact of different gates,
Fig.~\ref{fig:adder-probabilities}(c) shows an improvement of the
probability of obtaining the correct result when individual $\mathsf{FA}$'s
are processed individually one-by-one.

Whether dynamical bottlenecks can be completely eliminated in a proper
network of relaxed spins, possibly by the price of introducing additional
internal dynamical spins, ascends to the question whether the isotonicity
of a Boolean function is sufficient for ensuring single stable equilibrium
of the network. For the case of a scalar Boolean function without internal
spins, the answer is affirmative, as was shown above. Functions of a more
complex structure are a subject of ongoing research.

\section{Discussion}

We have demonstrated that classical dynamical systems can perform multiple
universal computations simultaneously within a single system, without
spatial or temporal separation of tasks, through the mechanism of
symmetry-induced parallelism, in which multiple inputs and their
corresponding outputs are contained within the symmetry of the dynamical
system. A distinguishing feature of this mechanism is that it holds for
nonlinear systems and therefore does not limit the character of
computations. We show that relaxed spin networks driven by the \mdII{}
model enable this mechanism owing to the model's dynamical properties.

A specific feature of the \mdII{} model is that its group of global
symmetries is one-parameter. This imposes an ordering on the relaxed spin
network phase space, and, as a result, simultaneous computation arises from
bit-flip chains of arguments, including the parameterization of the family
of evaluated functions, that are traversed by the symmetry transformation.
We show that simultaneous multiple evaluations correspond to the evaluated
function being isotone along the induced bit-flip chain. The isotonicity
ensures that there exists an association between input and output bits such
that the variation of the output bits simultaneous with the inversion of
the input bits along the induced chain is consistent with the function's
truth table. Viewed from the perspective of the relaxed spins phase space,
such an association presents a physical manifestation of symmetry-induced
parallelism. In turn, the role of the \mdII{} model is to dynamically solve
the combinatorial problem of associating the input and output relaxed spins
so as to form a state producing a valid chain of outputs under the symmetry
transformation. For a simple class of Boolean functions, we show that the
isotonicity of the evaluated function directly determines the dynamical
properties of the relaxed spin network and, consequently, guarantees the
convergence to the proper terminal state representing multiple
computations. Whether the isotonicity of Boolean functions with a more
complex structure ensures the convergence is an open problem. Resolving
this problem will determine whether symmetry-induced parallelism is a
deterministic feature that can be purposefully reproduced in spin
representations of elaborate circuits or whether it unavoidably becomes
probabilistic as a consequence of some inherent complexity of Boolean
functions.

The reliance on isotonicity may appear to be a \emph{coincidental}
consequence of the symmetry group being one-parameter.
At the same time, the principle of symmetry-induced parallelism by itself
does not immediately impose inherent limitations on the character of
simultaneous computations. It is therefore important to emphasize a
different perspective: isotonicity presents a foundation for computational
properties of the association between input and output bits. Such an
association, in turn, is a physical mechanism for synchronously traversing
various arguments and outputs by a symmetry transformation that acts
locally on the network graph and affects individual relaxed spins
independently. Thus, the limitations on the realization of symmetry-induced
parallelism considered in the present paper originate from the nature of
the symmetry transformations rather than from the symmetry group being
one-parameter.

\section{Methods}

\subsection{\mdII{} model}

The \mdII{} model is defined by the form of the objective function
characterizing the spin configuration. For example, the relaxed cut
function determining the evolution of the \mdII{} model can be written as
\begin{equation}\label{eq:v2-lyapunov}
  \mathcal{C}_{\IIind}(\boldsymbol{\sigma}, \mathbf{X}) =
  \mathcal{C}(\boldsymbol{\sigma}) + \Delta \mathcal{C}_{\IIind}(\boldsymbol{\sigma}, \mathbf{X}),
\end{equation}
where $\mathcal{C}(\boldsymbol{\sigma}) = \sum_{m,n} A_{m,n} \left( 1 - \sigma_{m} \sigma_{n}
\right)/4$ is the weight of the cut induced by
$\boldsymbol{\sigma}$ (the sign of $\sigma_{m}$ defines the partition assigned to
node $m$), and
\begin{equation}\label{eq:v2-cut-correction}
  \Delta \mathcal{C}_{\IIind}(\boldsymbol{\sigma}, \mathbf{X}) = \frac{1}{4}
  \sum_{m,n} A_{m,n} \sigma_{m} \sigma_{n} \abs{X_{m} - X_{n}}
\end{equation}
is the correction due to relaxation.

The equations of motion governing the evolution of the relaxed spins are
formulated as ensuring that
$\mathcal{C}_{\IIind}(\boldsymbol{\psi}) = \mathcal{C}_{\IIind}(\boldsymbol{\sigma}, \mathbf{X})$ does
not decrease with time:
$\dot{X}_{m} = 2 \partial \Delta \mathcal{C}_{\IIind}(\boldsymbol{\sigma}, \mathbf{X}) / \partial X_{m}$ (the
factor $2$ is introduced to simplify the expressions), or, more explicitly,
\begin{equation}\label{eq:v2-eqm}
  \frac{d X_{m}}{dt} =
  \sigma_{m} \sum_{n} A_{m,n} \sigma_{n} \sgn \left( X_{m} - X_{n} \right) .
\end{equation}
Here, $\sgn(X)$ is the sign function ($\sgn(X) = 1$, if $X>0$, and
$\sgn(X) = -1$, if $X < 0$) with the convention $\sgn(0) = 0$. Despite the
discontinuous variation of both components of the relaxed spin, the relaxed
cut function $\mathcal{C}_{\IIind}(\boldsymbol{\psi})$ and rates
$dX_{m} / dt$ remain continuous during these transitions.

\bibliography{logic.bib}

\section*{Acknowledgements}

The work was supported by the US National Science Foundation (NSF)
under Grant No. 2531175. The authors would like to thank Gitindra Sanyal
from the University of Michigan for useful discussions.

\end{document}